\documentclass{PoS}

\title{A perturbative study of the chirally rotated Schr\"{o}dinger Functional in QCD}
 
\ShortTitle{Perturbative study of the $\chi$SF in QCD}

\author{Stefan Sint\\
        School of Mathematics, Trinity College, Dublin 2, Ireland\\
        E-mail: \email{sint@maths.tcd.ie}}

\author{\speaker{Pol Vilaseca}\\
        Instituto Nazionale di Fisica Nucleare (INFN), Sezione di Roma, \\P.le A. Moro 2, I-00185 Roma, Italy\\
        E-mail: \email{pol.vilaseca.mainar@roma1.infn.it}}

\abstract{The chirally rotated Schr\"odinger functional ($\chi$SF) renders the mechanism of automatic $O(a)$ improvement 
compatible with the Schr\"odinger functional (SF) formulation. 
Here we report on the determination to 1-loop order in perturbation theory of the 
renormalization coefficients necessary to achieve automatic $O(a)$ improvement and the boundary 
improvement coefficients needed to eliminate the extra boundary $O(a)$ effects present in any SF formulation. 
After this is done, we perform a set of tests of automatic $O(a)$ improvement and of the universality between standard 
and chirally rotated SF formulations.}

\FullConference{The 32nd International Symposium on Lattice Field Theory,\\
		23-28 June, 2014\\
		Columbia University New York, NY}

\begin{document}

\section{Introduction}

Schr\"{o}dinger functional schemes \cite{LNWW1992_lattSF} have been successfully used in several 
renormalization problems in
lattice field theory. In this formulation, however, the presence of temporal boundaries
together with local boundary conditions for the fields are a source of extra cutoff effects. 
A theory regulated with Wilson fermions is hence affected by lattice artefacts
coming from the bulk and also by those originating at the boundaries.
These can be removed following Symanzik's improvement program by adding a set of counterterms
to the action in the bulk and at the boundaries. An appealing alternative to the standard formulation
of Wilson fermions in the SF is the recently proposed
chirally rotated Schr\"odinger functional ($\chi$SF), 
which implements the mechanism of (bulk) automatic $O(a)$ 
improvement \cite{S2011_lattSF}. 
In the continuum (and chiral) limit it is directly related to the standard SF
formulation via a chiral rotation of the fermion fields. The chirally rotated fields
satisfy modified boundary conditions which respect a version of chiral symmetry 
augmented with a flavour structure. In this situation, the argument for automatic $O(a)$
improvement can be invoked in terms of a rotated version of parity. Physical
observables are then only affected by $O(a^{2})$ discretization effects (provided that the
effects from the boundaries have been removed) without the need of introducing new
operators in the bulk.

Here we report on a perturbative 
1-loop calculation of the coefficients
necessary for the renormalization and the improvement of the theory in the $\chi$SF set-up. 
After this is done, we perform a
set of tests (also within perturbation theory) confirming 
the universality between the standard 
and chirally rotated set-ups, as well
as the mechanism of automatic $O(a)$ improvement.

\section{The $\chi$SF set-up}

For a flavour doublet chirally rotated boundary conditions take the form \cite{S2011_lattSF}
\begin{equation}
 \left.\widetilde{Q}_{+}\psi(x)\right|_{x_{0}=0} =  \left.\widetilde{Q}_{-}\psi(x)\right|_{x_{0}=T} =0,
\qquad \left.\overline{\psi}(x)\widetilde{Q}_{+}\right|_{x_{0}=0} = \left.\overline{\psi}(x)\widetilde{Q}_{-}\right|_{x_{0}=T},
\label{eq:boundary_conditions}
\end{equation}
with the projectors $\widetilde{Q}_{\pm}=\frac{1}{2}(1\pm i\gamma_{0}\gamma_{5}\tau^{3})$
and where $\tau^{i}$ are the Pauli matrices.

The projectors commute  with $i\gamma_{0}\gamma_{5}\tau^{3}$. A rotated version of
parity $P_{5}$ can hence be used to distinguish between even and odd observables
in the $\chi$SF\footnote{
$P_{5}:\psi(x)\rightarrow i\gamma_{0}\gamma_{5}\tau^{3}\psi(\widetilde{x}),\quad 
P_{5}:\overline{\psi}(x)\rightarrow -\overline{\psi}(\widetilde{x})i\gamma_{0}\gamma_{5}\tau^{3},\quad
\widetilde{x}=(x_{0},-{\bf x})$.}. 
On the lattice $P_{5}$-even correlations are automatically $O(a)$
improved and all $O(a)$ effects fall into $P_{5}$-odd observables.

The boundary conditions Eq.(\ref{eq:boundary_conditions}) can be derived
from the standard SF boundary conditions by applying a non-anomalous chiral
rotation to the flavour doublet
\begin{equation}
 \psi\rightarrow R(\alpha)\psi,\quad \overline{\psi}\rightarrow\overline{\psi}R(\alpha),
\quad R(\alpha)=\exp(i\alpha\gamma_{5}\tau^{3}/2).
\label{eq:rotation}
\end{equation}
The rotated fields satisfy Eq.(\ref{eq:boundary_conditions}) for $\alpha=\pi/2$
(standard SF boundary conditions are recovered for $\alpha=0$).
Since the rotation $R(\alpha)$ is a symmetry of the massless continuum
action, the two set-ups are equivalent, with correlation functions related through
\begin{equation}
 \langle\mathcal{O}[\psi,\overline{\psi}]\rangle_{\chi\textrm{\small SF}}=
 \langle\mathcal{O}[R(-\pi/2)\psi,\overline{\psi}R(-\pi/2)]\rangle_{\textrm{\small SF}}.
\label{eq:corr_rel}
\end{equation}

When the theory is implemented on the lattice with Wilson quarks, these 
relations are expected to hold between renormalized correlation functions
once the continuum limit is taken. Implementing $\chi$SF boundary conditions on the 
lattice is non-trivial. Here we consider the lattice set-up from \cite{S2011_lattSF} 
in which the fermionic action reads
\begin{equation}
 S_{f}=a^{4}\sum_{x_{0}=0}^{T}\sum_{\bf{x}}
\overline{\psi}(x)\left(\mathcal{D}_{W} + \delta\mathcal{D}_{W} + m_{0} \right)\psi(x),
\label{eq:XSF_action}
\end{equation}
with the Wilson-Dirac operator, which includes the clover term\footnote{Although the clover term is not needed for automatic $O(a)$ improvement,
including it removes some $O(a)$ effects from $P_{5}$-odd quantities.}, 
reads
\begin{equation}
 a \mathcal{D}_{W}\psi(x) = \left\{\begin{array}{ll}
    -U_{0}(x)P_{-}\psi(x+a\hat0)+(K+i\gamma_{5}\tau^{3}P_{-})\psi(x), & \quad x_{0}=0,\\
    -U_{0}(x)P_{-}\psi(x+a\hat0)+K\psi(x)-U_{0}(x-a\hat0)^{\dag}P_{+}\psi(x-a\hat0), & \quad 0<x_{0}<T,\\ 
(K+i\gamma_{5}\tau^{3}P_{+})\psi(x)-U_{0}(x-a\hat0)^{\dag}P_{+}\psi(x-a\hat0), &\quad x_{0}=T,
\end{array} \right.
\label{eq:XSF_WD_op}
\end{equation}
with the diagonal part $K$ given by
\begin{equation}
 K\psi(x)=\left(1+\frac{1}{2}\sum_{k=1}^{3}\left\{a\left(\nabla_{k}+\nabla_{k}^{*}\right)\gamma_{k}-a^{2}\nabla_{k}^{*}\nabla_{k}\right\}\right)\psi(x)
+c_{\textrm{{\tiny SW}}}\frac{i}{4}a\sum_{\mu,\nu=0}^{3}\sigma_{\mu\nu}\hat F_{\mu\nu}(x)\psi(x).
\label{eq:time_diag_K}
\end{equation}
The boundary counterterms read
\begin{equation}
 \delta \mathcal{D}_{W}\psi(x)=\left(\delta_{x_{0},0}+\delta_{x_{0},T}\right)\left[(z_{f}-1)+(d_{s}-1)a{\bf D}_{s}\right]\psi(x),
\label{eq:bound_count_XSF}
\end{equation}
with the operator
\begin{equation}
a{\bf D}_{s}=\frac{a}{2}\sum_{k}\left(\nabla_{k}+\nabla_{k}^{*}\right)\gamma_{k}-\frac{a^{2}}{2}\sum_{k}\nabla_{k}^{*}\nabla_{k}.
\label{eq:Ds} 
\end{equation}
Here $z_{f}$ is the coefficient of a dimension 3 boundary counterterm necessary
to restore the $P_{5}$ symmetry which is broken by the lattice regulator. 
The coefficient $d_{s}$ 
multiplies a dimension 4 counterterm at the boundaries and can be tuned to remove
$O(a)$ effects from the boundaries. While $d_{s}$ can be understood as the 
$\chi$SF counterpart of the $\widetilde{c}_{t}$ coefficient in the SF, $z_{f}$
is special from this set-up \cite{S2011_lattSF}. 
In perturbation theory these coefficients read
\begin{equation}
 z_{f}=z_{f}^{(0)} + g_{0}^{2}z_{f}^{(1)} + O(g_{0}^{4}),\qquad
 d_{s}=d_{s}^{(0)} + g_{0}^{2}d_{s}^{(1)} + O(g_{0}^{4}),
\end{equation}
where $z_{f}^{(0)}=1$ and $d_{s}^{(0)}=1/2$. One of the central goals of this work
is to determine $z_{f}^{(1)}$ and $d_{s}^{(1)}$. Although knowing $d_{s}$ to
1-loop is enough in practise, in a non-perturbative calculation $z_{f}$ must
be known also non-perturbatively in order to ensure that the correct symmetries 
are recovered in the continuum limit \cite{DB2014}. The knowledge of
$z_{f}^{(1)}$ can help to guide the non-perturbtive tuning and it is moreover
required in further perturbative calculations.
\section{Correlation functions in the SF and the $\chi$SF}
\label{sec:corrs}
Correlation functions in the $\chi$SF set-up are defined in a similar way as 
in the standard SF. 
Writing explicitly the flavour assignments, boundary to 
bulk correlation functions are given by
\begin{equation}
 \textrm{g}^{f_{1}f_{2}}_{X}(x_{0})= -\frac{1}{2}\langle X^{f_{1}f_{2}}(x_{0})\mathcal{Q}_{5}^{f_{2}f_{1}}\rangle,\qquad 
\textrm{and} \qquad
 l^{f_{1}f_{2}}_{Y}(x_{0})= -\frac{1}{6}\sum_{k=1}^{3}\langle Y_{k}^{f_{1}f_{2}}(x_{0})\mathcal{Q}_{k}^{f_{2}f_{1}}\rangle,
\label{eq:gl_corr}
\end{equation}
with the fermion bilinears being $X=A_{0},V_{0},S,P,$ and $Y_{k}=A_{k},V_{k},T_{k0},\widetilde{T}_{k0}$.
Boundary to boundary correlation functions read
\begin{equation}
 \textrm{g}_{1}^{f_{1}f_{2}}=-\frac{1}{2}\langle \mathcal{Q}_{5}^{f_{1}f_{2}}\mathcal{Q'}_{5}^{f_{2}f_{1}}\rangle,
\qquad\textrm{and}\qquad
 l_{1}^{f_{1}f_{2}}=-\frac{1}{6}\sum_{k=1}^{3}\langle \mathcal{Q}_{k}^{f_{1}f_{2}}\mathcal{Q'}_{k}^{f_{2}f_{1}}\rangle.
\label{eq:g1_corr}
\end{equation}
The boundary operators $\mathcal{Q}_{5}^{f_{1}f_{2}},...,$ are constructed by applying 
the rotation Eq.(\ref{eq:rotation}) to the standard SF boundary operators such 
that Eq.(\ref{eq:corr_rel}) holds.
For avoiding the computation of correlation functions of flavour singlets 
we consider a set-up with 2 types of up quarks (u and u') and 2 types of 
down quarks (d and d') \cite{L2010}.


For the standard SF we use correlation functions following conventions from 
the literature.
Following Eq.(\ref{eq:corr_rel}) we can write a dictionary relating correlation
functions in the 2 set-ups similar to that relating standard 
and twisted mass QCD. Some relations with the $P_{5}$-even correlations are 
\begin{equation}
 f_{A}=\textrm{g}_{A}^{uu'}=-i\textrm{g}_{V}^{ud},\qquad f_{P}=i\textrm{g}_{S}^{uu'}=\textrm{g}_{P}^{ud},    \\
 \qquad k_{V}=l_{V}^{uu'}=-il_{A}^{ud},\qquad  k_{T}=il_{\widetilde{T}}^{uu'}=l_{T}^{ud},
\label{f_g_even}
\end{equation}
and with $P_{5}$-odd correlation functions
\begin{equation}
 f_{V}=\textrm{g}_{V}^{uu'}=-i\textrm{g}_{A}^{ud},\qquad f_{S}=i\textrm{g}_{P}^{uu'}=\textrm{g}_{S}^{ud}, \\
 \qquad k_{A}=l_{A}^{uu'}=-il_{V}^{ud},\qquad k_{\widetilde{T}}=il_{T}^{uu'}=l_{\widetilde{T}}^{ud}.
\label{f_g_odd}
\end{equation}

The tuning of $z_{f}$ and $m_{c}$ accounts for the restoration of $P_{5}$
and chiral symmetries and must be done simultaneously. A typical condition
for fixing $m_{0}=m_{c}$ is to demand the PCAC mass to be zero at the middle
of the lattice. For determining $z_{f}$ one can require any $P_{5}$-odd
quantity to vanish. In this study we consider 4 renormalization conditions
for $z_{f}$, i.e.
\begin{equation}
 i)~\textrm{g}_{P}^{uu'}=0,\qquad ii)~\textrm{g}_{A}^{ud}=0, \qquad 
 iii)~\textrm{g}_{V}^{uu'}=0, \qquad \textrm{and} \qquad iv)~\textrm{g}_{S}^{ud}=0.
\label{eq:renconds}
\end{equation}
Different renormalization conditions for $z_{f}$ lead to differences 
$\Delta z_{f}$ which vanish linearly in $a/L$.
\section{Perturbation theory}
In perturbation theory, the correlation functions of the previous
subsection are expanded to 1-loop order as
\begin{equation}
\textrm{g}_{X}(x_{0}) = \textrm{g}_{X}^{(0)}(x_{0}) + \textrm{g}_{X}^{(1)}(x_{0})g_{0}^{2} + O(g_{0}^{4}), \qquad
\textrm{g}_{1} = \textrm{g}_{1}^{(0)} + \textrm{g}_{1}^{(1)}g_{0}^{2} + O(g_{0}^{4}) ,
\end{equation}
and similarly with all the other correlation functions. The gauge fixing
procedure is the same as in \cite{L1996}, and so is the gluon propagator. The 
calculation of $\textrm{g}_{X}^{(1)}$ and $\textrm{g}_{1}^{(1)}$
requires de evaluation of the same set of diagrams as those
shown in \cite{L1996,S1997}, together with the contribution due to the $\chi$SF
boundary counterterms Eq.(\ref{eq:bound_count_XSF}).  Explicit expressions for the vertices and quark
propagator derived from Eq.(\ref{eq:XSF_WD_op}) will be 
given elsewhere \cite{V2014}.

We have produced a program for the fast evaluation of a large set 
of correlation functions of fermion bilinears to 1-loop in perturbation
theory for both the standard and chirally rotated set-ups.
We calculate correlation functions for $L/a\in[6,48]$ from which it is possible to extract
the different terms of the asymptotic expansion of the 1-loop coefficients.
\subsection{Determination of $m_{c}^{(1)}$ and $z_{f}^{(1)}$}
The 1-loop coefficients $m_{c}^{(1)}$ and $z_{f}^{(1)}$ are obtained
by expanding the renormalization conditions in section \ref{sec:corrs}
and solving them up to 1-loop order for a range of lattice spacings.
After taking the continuum extrapolation $a/L\rightarrow0$ we obtain
\begin{equation}
 \left\{\begin{array}{l}
m_{c}^{(1)}(c_{\textrm{{\tiny SW}}}=1)=-0.2025565(1)\times C_{2}(R),\\
m_{c}^{(1)}(c_{\textrm{{\tiny SW}}}=0)=-0.325721(7)\times C_{2}(R),
         \end{array}\right.\qquad
 \left\{\begin{array}{l}
z_{f}^{(1)}(c_{\textrm{{\tiny SW}}}=1)=0.167572(2)\times C_{2}(R),\\
z_{f}^{(1)}(c_{\textrm{{\tiny SW}}}=0)=0.33023(6)\times C_{2}(R).
         \end{array}\right.
\label{mc1_value}
\end{equation}
where $C_{2}(R)$ is the quadratic casimir operator in the representation\footnote{$C_{2}(F)=(N^{2}-1)/2N$ for the fundamental representation.}
$R$.

The coefficient $m_{c}^{(1)}$ reproduces the known values of the critical
mass for $c_{\textrm{{\tiny SW}}}=1$ and $0$, as expected. 
The coefficient $z_{f}^{(1)}$ has been calculated
here for the first time. It is worth noting that the determination
of $m_{c}$ is quite independent from $z_{f}$, which has also been 
observed in quenched calculations \cite{L2010,GL2013} and in dynamical 
studies \cite{DB2014}.

\begin{figure}[!ht]
\centering
\includegraphics[clip=true,scale=0.48]{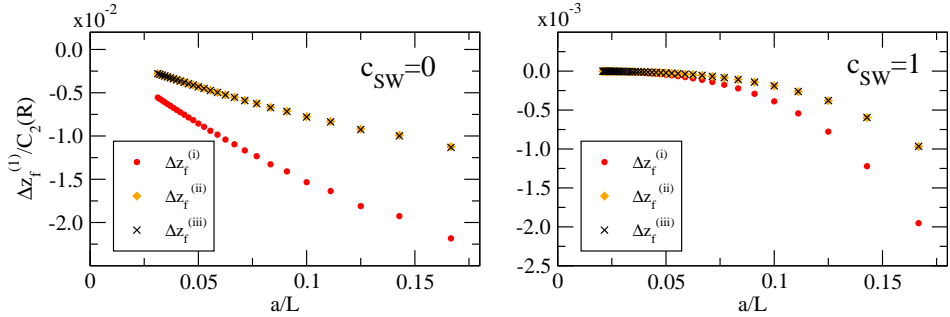}
\caption{Differences in the 1-loop value $z_{f}^{(1)}$ 
at finite lattice spacing for the different tuning
conditions.}
\label{fig:zf_conditions}
\end{figure}

Next we calculate 
the differences $\Delta z_{f}^{(X)}=\left.z_{f}^{(1)}\right|_{X}-\left.z_{f}^{(1)}\right|_{iv}$ 
in determining $z_{f}$ using the different renormalization conditions in Eq.(\ref{eq:renconds}).
In figure \ref{fig:zf_conditions} it can be seen that $\Delta z_{f}^{(X)}$ vanish linearly as
$a/L\rightarrow0$ for $c_{\textrm{{\tiny SW}}}=0$, while for $c_{\textrm{{\tiny SW}}}=1$ the convergence is much
faster.
\subsection{Determination of $d_{s}^{(1)}$}
The determination of the 1-loop boundary improvement
coefficient $d_{s}^{(1)}$ can be done by requiring
the absence of $O(a)$ terms at 1-loop in some 
$P_{5}$-even quantity. 
We consider an even quantity evaluated to 1-loop order and at several values of the fermion
boundary angle $\theta$, i.e.
\begin{equation}
\left.\frac{\left[\textrm{g}_{P}^{ud}(x_{0},\theta,a/L)\right]_{R}}{\left[\textrm{g}_{P}^{ud}(x_{0},0,a/L)\right]_{R}}\right|_{x_{0}=T/2},
\qquad d_{s}^{(1)}=-0.0009(3)\times C_{2}(R).
\label{ratio_XSF}
\end{equation} 
For $\theta=0.1$, $0.5$ and $1.0$ we consistently find the value of $d_{s}^{(1)}$ given in Eq.(\ref{ratio_XSF}).
\subsection{Test of automatic $O(a)$ improvement}
Once the determination of the improvement and renormalization coefficients has 
been done we would like to test whether
the mechanism of automatic $O(a)$ improvement works.

\begin{figure}[!ht]
\centering
\includegraphics[clip=true,scale=0.50]{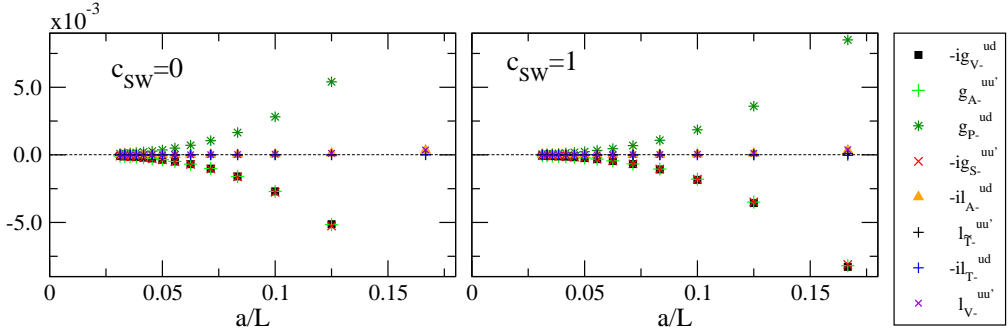}
\caption{Vanishing correlation functions with reversed projectors in the boundary operators.}
\label{fig:even_van}
\end{figure}

First of all we check that the boundary conditions Eq.(\ref{eq:boundary_conditions})
are correctly realised. After $z_{f}$ is tuned to its critical value 
eq.(\ref{eq:boundary_conditions}) should hold up to cutoff effects. To test
this we evaluate a set of even correlation functions for which the projectors
at the boundary operators have been reverted\footnote{These correlations are labelled
with a ``$-$``sign, i.e. $g_{X,-}^{f_{1}f_{2}}$.} $Q_{\pm}\rightarrow Q_{\mp}$.
Such correlations should vanish provided that boundary conditions are correctly
implemented. In figure \ref{fig:even_van} it can be seen that correlation functions
with reverted projectors are very small and vanish as the continuum limit is approached. 

\begin{figure}[!t]
\centering
\includegraphics[clip=true,scale=0.50]{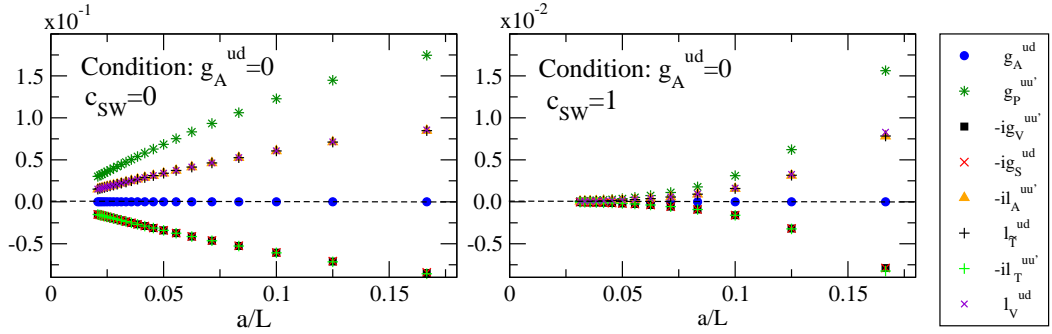}
\caption{Vanishing $P_{5}$-odd correlation functions at 1-loop calculated for $c_{\textrm{{\tiny SW}}}=0$ and $1$, 
 with $z_{f}$ fixed using the renormalization condition $\textrm{g}_{A}^{ud}=0$.}
\label{fig:odd_van_1}
\end{figure}

Secondly, we study the continuum limit of $P_{5}$-odd correlation functions and verify
that these, being pure cutoff effects, vanish with the expected rate in $a/L$ (see figure \ref{fig:odd_van_1}).
\subsection{Universality}
\begin{figure}[!t]
\centering
\includegraphics[clip=true,scale=0.50]{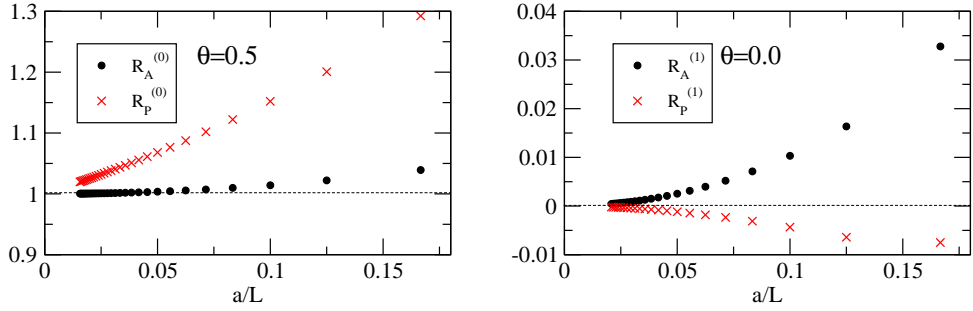}
\caption{Ratios $R_{A}$ and $R_{P}$ at tree level (left panel) and 1-loop (right panel).
For the tree level ratios, the fermion boundary angle is chosen to be $\theta\neq0$.
Otherwise, for the choice $\theta=0$ cutoff effects are absent from tree-level correlation functions
and the ratio is exactly 1.}
\label{fig:univ}
\end{figure}

Once the coefficients $z_{f}$ and $m_{c}$ are known, the equalities in Eqs.(\ref{f_g_even}) and (\ref{f_g_odd}) are expected to hold between ratios
of renormalized correlation functions. For instance, the ratios
\begin{equation}
 R_{A} = \left[\frac{\textrm{g}_{A}^{uu'}(T/2)}{\sqrt{\textrm{g}_{1}^{uu'}}}\right]\times\left[\frac{f_{A}(T/2)}{\sqrt{f_{1}}}\right]^{-1},
 \qquad \textrm{and}\qquad
 R_{P} = \left[\frac{\textrm{g}_{P}^{ud}(T/2)}{\sqrt{\textrm{g}_{1}^{ud}}}\right]\times\left[\frac{f_{P}(T/2)}{\sqrt{f_{1}}}\right]^{-1},
\end{equation}
should approach $1$ as $a/L\rightarrow0$. In perturbation theory, the ratios are expanded as
\begin{equation}
 R_{X}=R_{X}^{(0)}+g_{0}^{2}R_{X}^{(1)}+O(g_{0}^{4}).
\end{equation}
Universality implies that $R_{X}^{(0)}\rightarrow1$ and $R_{X}^{(1)}\rightarrow0$ as $a/L\rightarrow0$. As can be seen from
figure \ref{fig:univ}, we confirm the expected convergence of the universality relations as the continuum limit is approached.
\vspace{-0.5cm}
\section{Conclusions}
Here we have calculated for the first time to 1-loop order in perturbation
theory the renormalization and $O(a)$ improvement coefficients 
$z_{f}^{(1)}$ and $d_{s}^{(1)}$ for the $\chi$SF set-up. With the knowledge of
these coefficients we have confirmed, always in the framework of perturbation
theory, that automatic $O(a)$ improvement is at work. Also, the universality
between standard and chirally rotated frameworks is confirmed. 
The $\chi$SF opens new possibilities for determining finite renormalization
constants and $O(a)$ improvement coefficients for theories with Wilson type
fermions \cite{DB2014}. Further perturbative calculations are
essential to determine them in a way in which cutoff effects are minimal.
\section*{Acknowledgements}
This work received funding from the Research Executive
Agency (REA) of the European Union under Grant Agreement number PITN-GA-2009-238353
(ITN STRONGnet). SS also acknowledges support by SFI under grant 11/RFP/PHY3218.

\end{document}